\documentclass{ws-mplb}

\begin{document}

\markboth{Danh-Tai HOANG, Yann MAGNIN, H. T. DIEP}{Spin Resistivity in the Frustrated $J_1-J_2$ Model}

\catchline{}{}{}{}{}
%
%%%%%%%%%%%%%%%%%%%%%%%%%%%%%%%%%%%%%%%%%%%%%%%%%%%%%%%%%%%%%%%%%%%%

\title{SPIN RESISTIVITY IN THE FRUSTRATED $J_1-J_2$ MODEL}

\author{\footnotesize Danh-Tai HOANG\footnote{danh-tai.hoang@u-cergy.fr}, Yann MAGNIN\footnote{yann.magnin@u-cergy.fr} and H. T. DIEP\footnote{diep@u-cergy.fr} \footnote{Corresponding author}}

\address{Laboratoire de Physique Th\'eorique et Mod\'elisation,
Universit\'e de Cergy-Pontoise,\\ CNRS UMR 8089\\
2, Avenue Adolphe Chauvin, 95302 Cergy-Pontoise Cedex, France.}

\maketitle

\begin{history}
\received{(Day Month Year)}
\revised{(Day Month Year)}
\end{history}

\begin{abstract}
We study in this paper the resistivity encountered by Ising itinerant spins
traveling in the so-called $J_1-J_2$ frustrated simple cubic Ising lattice.  For the lattice, we take into account the interactions between
nearest-neighbors and next-nearest-neighbors,  $J_1$ and  $J_2$ respectively.  Itinerant spins interact with lattice spins via a distance-dependent interaction.  We also take into account an interaction between itinerant spins.
The lattice is frustrated in a range of $J_2$ in which we show that it undergoes a very strong first-order transition.   Using Monte Carlo simulation, we calculate the resistivity $\rho$ of the itinerant spins and show that the first-order transition of the lattice causes a discontinuity of $\rho$.
\end{abstract}


\section{Introduction}

One of the most fascinating subjects in condensed matter physics is the study of the resistivity encountered by conduction electrons in crystals.  Fifty years ago, the effect of the magnetic ordering of the crystal on the electron resistivity began to attract investigations.  It has been shown that at very low temperature ($T$) the resistivity is dominated by spin-wave scattering, the spin resistivity $\rho$ is then proportional to $T^2$ in ferromagnets.\cite{Kasuya,Turov}  In the region of the ferromagnetic phase transition,  $\rho$ shows a peak at the transition temperature $T_C$ similar to the magnetic susceptibility. De Gennes and Friedel\cite{DeGennes} have suggested that this behavior is due to the spin-spin correlation.  Several approximations have been used to treat this correlation appearing  in several formulations,\cite{Fisher,Zarand} in particular in the Boltzmann's equation.\cite{Haas,Kataoka}  Recently, we have introduced a Monte Carlo (MC) simulation technique to deal with the spin resistivity.  Our results for ferromagnets are in agreement with other theories, in particular the existence of the peak at $T_C$ and its dependence on the strength of magnetic field and density of itinerant spins.\cite{Akabli,Akabli2}  In unfrustrated antiferromagnets,  there has been a few investigations. Some theories predicted the absence of a sharp peak at $T_C$.\cite{Haas}  We have shown by MC simulations that this is true, however the form of the rounded peak depends on the crystal structure and other interaction parameters.\cite{Akabli3,Magnin}

Experimentally, there has been a large number of works dealing with the spin resistivity in different magnetic compounds.\cite{Stishov,Stishov2,Chandra,Xia,Wang-Chen,Santos,Li,Lu,Du,Zhang,McGuire}  These experiments show the existence of an anomaly of $\rho$ at the magnetic phase transition. The shape of this anomaly depends on the material.

One class of interesting magnetic systems is called "frustrated systems" introduced in the 1970s in the context of spin glasses.   These frustrated systems are very unstable due to the competition between different kinds of interaction. However they are periodically defined (no disorder) and therefore subject to exact treatments.  This is the case of several models in two dimensions\cite{Diep-Giacomini}, but in three dimensions frustrated systems are far from being understood even on basic properties such as the order of the phase transition (first or second order, values of critical exponents, ...).   Let us recall the definition of a frustrated system.
When a spin cannot fully satisfy energetically all the interactions with its neighbors, it is
"frustrated".   This occurs when the interactions are in competition with
each other or when the lattice geometry does not allow to satisfy all interaction bonds simultaneously. A well-known example is the stacked triangular lattice with antiferromagnetic interaction between nearest-neighbors.
The frustration
in spin systems causes many unusual properties such as large ground state (GS)
degeneracy, successive phase transitions
with complicated nature, partially disordered phases, reentrance and disorder lines.
Frustrated systems still constitute at present a challenge for investigation methods.  For  recent reviews, the
reader is referred to Ref. \cite{Diep2005}.

 Motivated by their exotic behaviors, we have studied some frustrated systems and found that $\rho$ depends drastically on the range of interaction, and that $\rho$ shows a discontinuity at $T_C$ reflecting the first-order character of the phase transition.\cite{Magnin2}

This work aims at  confirming the fact that in systems with first-order transitions, $\rho$ should have a discontinuity at $T_C$.  For that purpose, we consider in this paper so-called $J_1-J_2$ simple cubic lattice with Ising spins.  This system is known to undergo a very strong first-order transition in the Heisenberg case.\cite{Pinettes}  The Ising case studied here shows also a very strong first-order transition as shown below.

In section 2, we present our model and MC method. The results are shown in section 3. Concluding remarks are given in section 4.

\section{Model and Method}

\subsection{Model}
\begin{figure}[th]
\vspace*{8pt}
\centerline{\epsfig{file=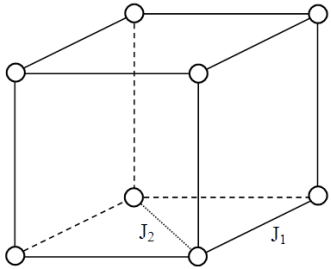,width=1.8in}}
\caption{Simple cubic lattice with nearest and next-nearest neighbor interactions, $J_1$ and $J_2$, indicated.} \label{fig:model}
\end{figure}

\begin{figure}[th]
\vspace*{8pt}
\centerline{\epsfig{file=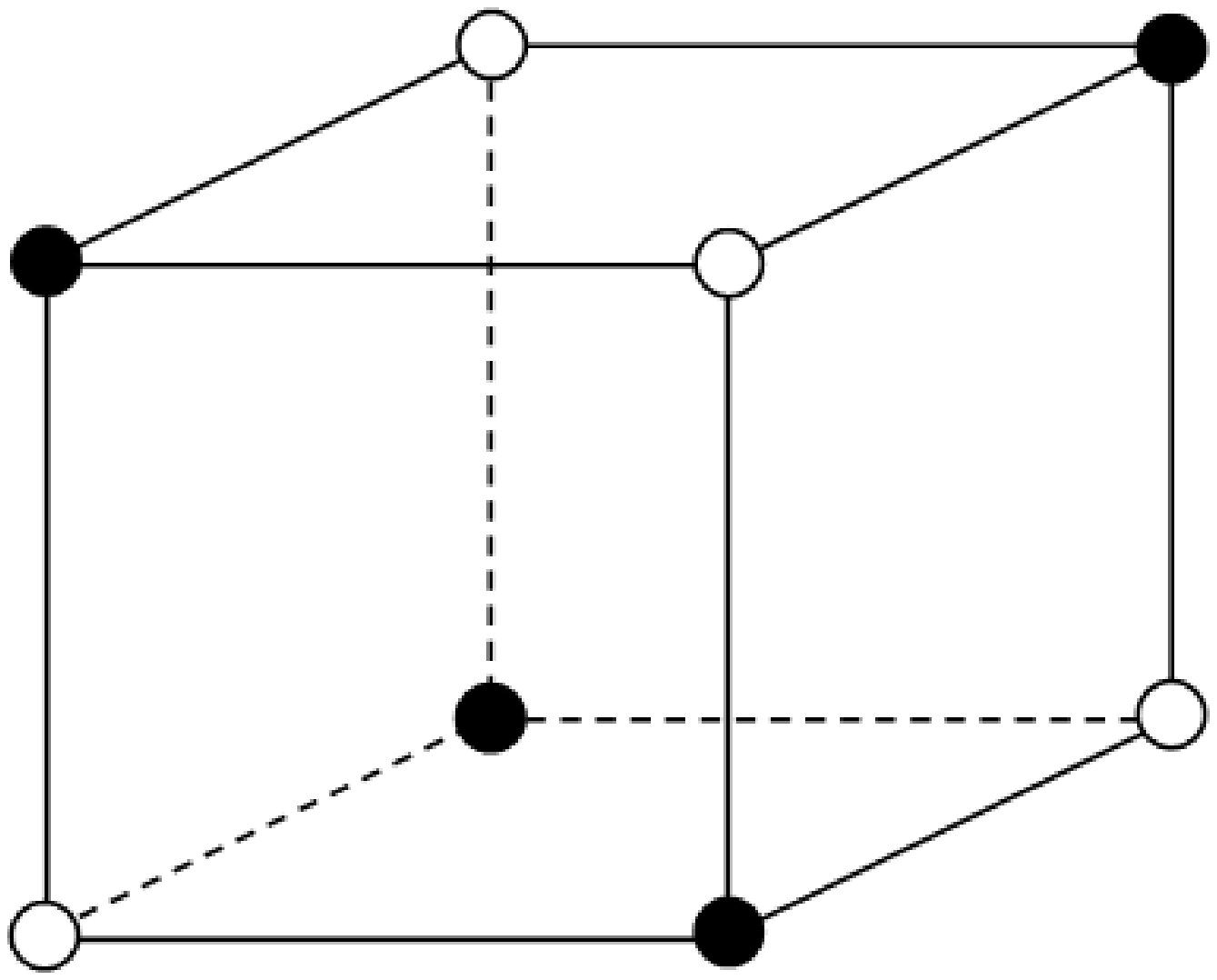,width=1.8in}}
\centerline{\epsfig{file=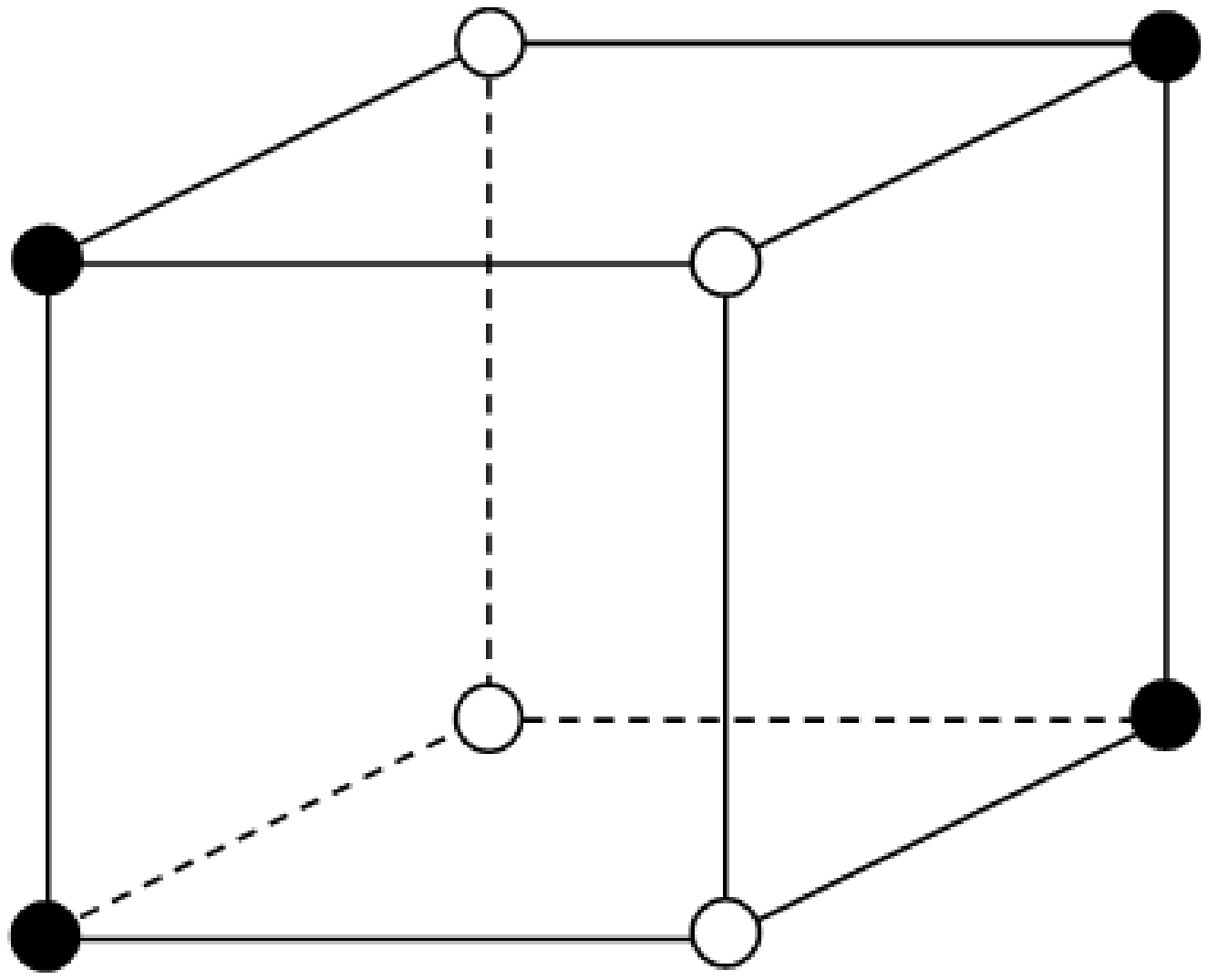,width=1.8in}}
\caption{Simple cubic lattice. Up-spins: white circles, down-spins: black circles. Upper: Ground state when $|J_2|<0.25 |J_1|$, lower: Ground state when $|J_2|>0.25 |J_1|$.} \label{fig:SC}
\end{figure}

We consider the simple cubic lattice shown  in Fig. \ref{fig:model}. The spins are
the classical Ising model of magnitude
$S=1$. The Hamiltonian is given by
\begin{equation}\label{HL}
{\cal H} = -J_1\sum_{(i,j)} \mathbf{S}_i.\mathbf{S}_j -J_2\sum_{(i,m)} \mathbf{S}_i.\mathbf{S}_m
\end{equation}
where $\mathbf{S}_i$ is the Ising spin at the lattice site $i$, $\sum_{(i,j)}$ is made
over the NN spin pairs $\mathbf{S}_i$ and $\mathbf{S}_j$ with  interaction $J_1$, while $\sum_{(i,m)}$ is performed over the NNN pairs with interaction $J_2$.  We are interested in the frustrated regime.  Therefore, hereafter we suppose that $J_1=-J$ ($J>0$, antiferromagnetic interaction, and $J_2=-\eta J$ where $\eta$ is a positive parameter.  The ground state (GS) of this system is easy to obtain either by minimizing the energy, or by comparing the energies of different spin configurations, or just a numerical minimizing by a steepest descent method.\cite{Ngo2007}  We obtain the antiferromagnetic (AF) configuration shown by the upper figure of  Fig. \ref{fig:SC} for $|J_2|<0.25 |J_1|$, or the configuration shown in the lower figure for $|J_2|>0.25 |J_1|$.  Note that this latter configuration is 3-fold degenerate by putting the parallel NN spins on $x$, $y$ or $z$ axis. With the permutation of black and white spins, the total degeneracy is thus 6.

The phase transition of this model in the frustrated region ($|J_2|>0.25 |J_1|$) has been studied by Pinettes and Diep\cite{Pinettes} in the case of the Heisenberg model.  It has been found that the transition is strongly of first order.  The ordered phase is very unstable due to its degeneracy. As will be shown below, the case of the Ising case shows an even stronger first-order transition.  It is therefore interesting to  investigate the resistivity of itinerant spins traveling across of such a system.

%\begin{figure}[th]
%\centerline{\psfig{file=mplbf1.eps,width=5cm}}
%\vspace*{8pt}
%\caption{A schematic illustration of dissociative
%recombination. The direct mechanism,
%4m$^2_\pi$ is initiated when the molecular ion S$_{\rm L}$ captures an
%electron with kinetic energy.\label{f1}}
%\end{figure}

The model we use here to study the spin transport takes into account the following interactions:
\begin{itemize}
\item Interaction between lattice spins given by Eq. (\ref{HL})

\item Interaction between itinerant spins and lattice spins given by
\begin{eqnarray}\label{HIL}
\mathcal{H}_r & = & -\sum_{i,j}I_{i,j}\vec{\sigma}_i .\vec{S}_j
\end{eqnarray}
where $\sigma_i$ is the Ising spin of itinerant electron and $I_{i,j}$ denotes the interaction that depends on the distance between an electron $i$ and the spin $\vec{S}_j$ at the lattice site $j$. We use the following interaction expression :
\begin{eqnarray}
I_{i,j} & = & I_{0}e^{-\alpha r_{ij}} \mbox{\hspace{0,3cm}with\hspace{0,3cm}} r_{ij}=|\vec{r}_i-\vec{r}_j|
\end{eqnarray}
where $I_0$ and $\alpha$ are  constants which will be chosen in section \ref{choice}.

\item Interaction between itinerant spins:
 In the same way, interaction between itinerant electrons is defined by
\begin{eqnarray}
\mathcal{H}_m & = & -\sum_{i,j}K_{i,j}\vec{\sigma}_i .\vec{\sigma}_j\label{HI}\\
K_{i,j} & = & K_{0}e^{-\beta r_{ij}}\label{K}
\end{eqnarray}
with $\sigma_i$ the spin of itinerant electron and $K_{i,j}$ the interaction that depends on the distance between electrons $i$ and $j$.  The choice of the constants $K_0$ and $\beta$ is discussed in \ref{choice}.

\item Chemical potential term:
Since the interaction between itinerant electron spins is attractive,  we need to add a chemical potential in order to avoid a possible collapse of electrons into some points in the crystal and to ensure a homogeneous distribution of electrons during the simulation. The chemical potential term is given by
\begin{eqnarray}
\mathcal{H}_c & = & D\vec{\nabla}_rn(\vec{r})\label{pot}
\end{eqnarray}
where $n(\vec r)$ is the concentration of itinerant spins in the sphere of $D_2$ radius, centered at $\vec r$. $D$ is a constant parameter appropriately chosen.

\item Electric field term:
\begin{eqnarray}
\mathcal{H}_E & = & -e\vec{\epsilon}.\vec{\ell}
\end{eqnarray}
where $e$ is the charge of electron, $\vec \epsilon $ the applied electrical field and $\vec \ell$ the displacement vector of an electron.\\
\end{itemize}

\subsection{Choice of parameters}\label{choice}
Note that the effect of the crystal magnetic ordering on the resistivity is dominated by the first two interactions.
We will show below results obtained for typical values of parameters.  The choice of the parameters has been made after numerous test runs.  We describe the principal requirements which guide the choice:

i) We choose the interaction between lattice spins as unity, i. e. $|J|=1$.

ii) We choose interaction between an itinerant and its surrounding lattice spins so as its energy $E_i$ in the low $T$ region is the same order of magnitude with that between lattice spins. To simplify, we take $\alpha=1$.

iii) Interaction between itinerant spins is chosen so that this contribution to the itinerant spin energy is smaller than
$E_i$ in order to highlight the effect of the lattice ordering on the spin current. To simplify, we take $\beta=1$.

iv) The choice of $D$ is made in such a way to avoid the formation of  clusters of itinerant spins (collapse) due to their attractive interaction [Eq. (\ref{K})].

v) The electric field is chosen not so strong in order to avoid its dominant effect that would mask the effects of thermal fluctuations and of the magnetic ordering.

vi) The density of the itinerant spins is chosen in a way that the contribution of interactions between themselves is neither so weak nor so strong with respect to $E_i$.

Within these requirements, a variation of each parameter does not change qualitatively the results shown below. As will be seen, only the variation of $D_1$ does change drastically the results. That is the reason why we will study in detail the effect of this parameter.  For larger densities of itinerant spins, the resistivity is larger as expected because of additional scattering process between itinerant spins.

We fix $J_1=-J=-1$ (AF interaction) for NN coupling of lattice spins as said above.  The energy is thus measured in the unit of $J$. The temperature is expressed in the unit of $J/k_B$.  The distance ($D_1$, $D_2$) is in the unit of $a$, the lattice constant.

\subsection{Simulation Method}

We consider a film with a thickness of $N_z$ cubic cells in the $z$ direction. Each of the $xy$ planes contains  $N_x\times N_y$ cells. The periodic boundary conditions are used on the $xy$ planes to ensure that the itinerant electrons who leave the system at the second end are to be reinserted at the first end. For the $z$ direction, we use the mirror reflection at the two surfaces.  These boundary conditions  conserve thus the average density of itinerant electrons.
Dynamics of itinerant electrons is created by an electric field applied along the $x$ axis.

Simulations are carried out in the following manner. The lattice spins are equilibrated at a temperature $T$. Itinerant spins are then injected into the system.  Before calculating thermal averages of transport properties, we equilibrate itinerant spins during a large number of MC steps.  The multi-step averaging procedure has been used to get good statistics:\cite{Magnin}  Averaging is made between re-equilibrating periods of lattice and itinerant spins to explore a maximum number of microscopic spin configurations.

\section{Results}

\begin{figure}[th]
\vspace*{8pt}
\centerline{\epsfig{file=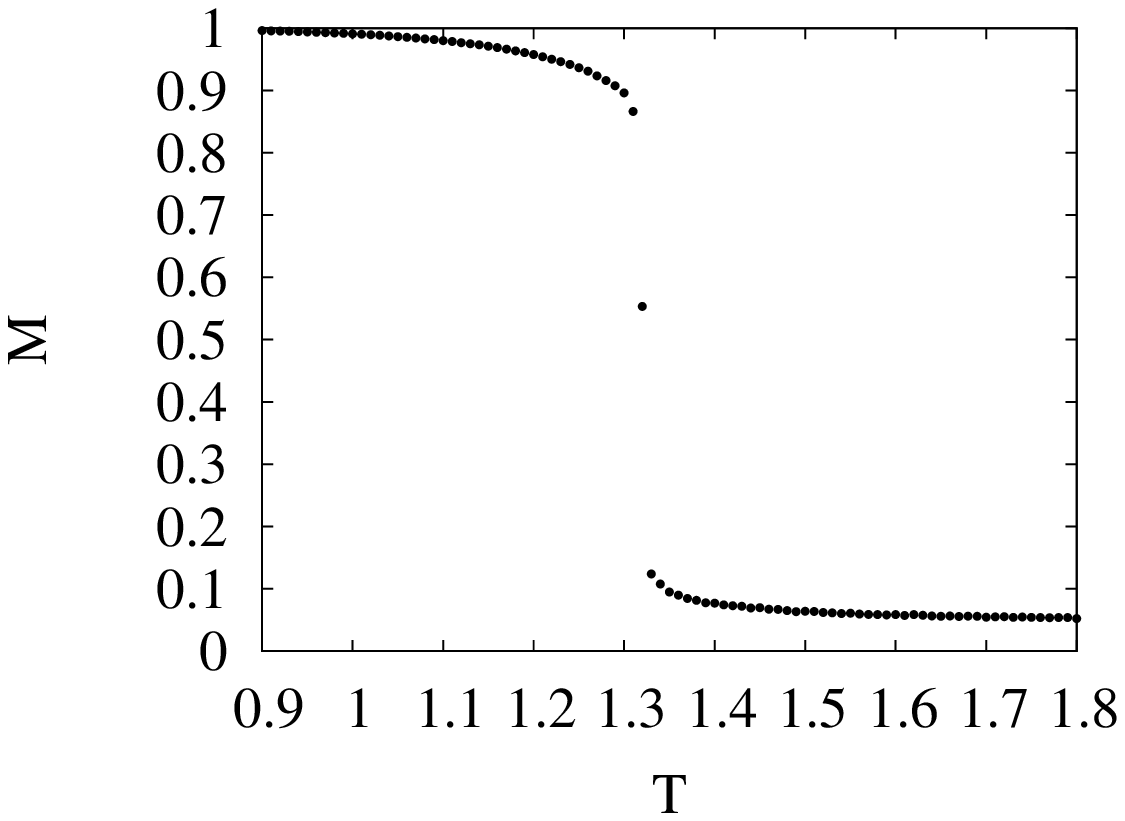,angle=0,width=2.3in}}
\centerline{\epsfig{file=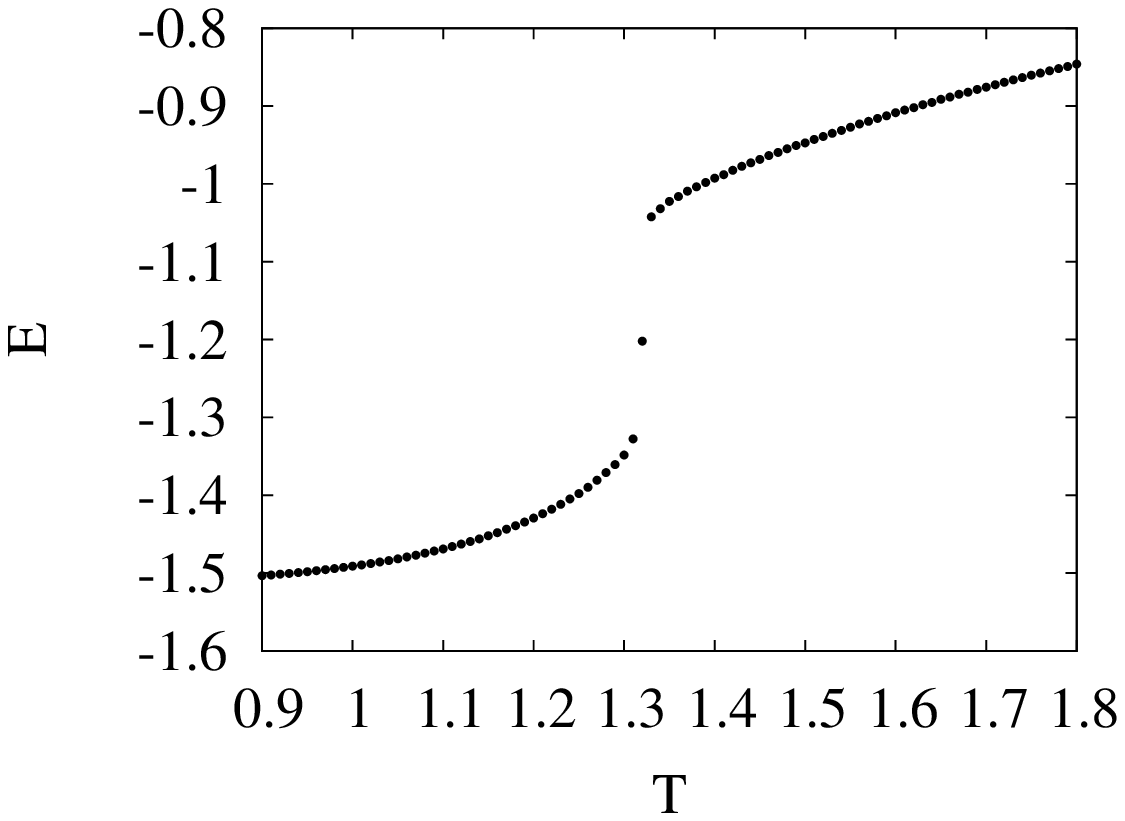,angle=0,width=2.3in}}
\caption{Upper: Sublattice magnetization $M$ versus $T$, Lower: Energy versus $T$,  for $|J_2|=0.26|J_1|$, $N_x=N_y=20$, $N_z=6$.} \label{fig:ME}
\end{figure}

\begin{figure}[th]
\vspace*{8pt}
\centerline{\epsfig{file=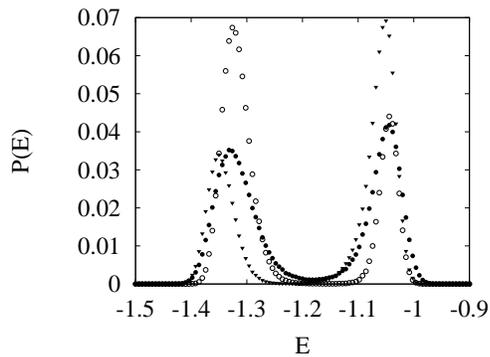,width=2.8in}}
\caption{Energy histogram taken at the transition temperature $T_C$ for $J_2=-0.26|J_1|$: black circles are for $N_x=N_y=20, N_z = 6, T_C=1.320$, void circles for $N_x=N_y=30, N_z = 6,T_C=1.320$ and black triangles for $N_x=N_y=20, N_z = 10, T_C=1.305$.  Other parameters are $I_0=K_0=0.5$, $D_1=0.8a$, $D_2=a$, $D=1$,  $\epsilon=1$.} \label{fig:HE}
\end{figure}

We show first the result of the lattice alone, namely without itinerant spins.  The lattice in the frustrated region, i. e. $|J_2/J_1|>0.25$, shows a strong first-order transition as seen in Fig. \ref{fig:ME}:  The  sublattice magnetization and the energy per spin as functions of $T$ for $J_2=-0.26|J_1|$ for the lattice size $N_x=N_y=20$, $N_z=6$ show a discontinuity at the transition temperature.  To check further the first-order nature of the transition, we have calculated the energy histogram at the transition temperature $T_C$.  This is shown in Fig. \ref{fig:HE}.  The double-peak structure indicates the coexistence of the ordered and disordered phases at $T_C$. The distance between two peaks represents the latent heat.

\begin{figure}[th]
\vspace*{8pt}
\centerline{\epsfig{file=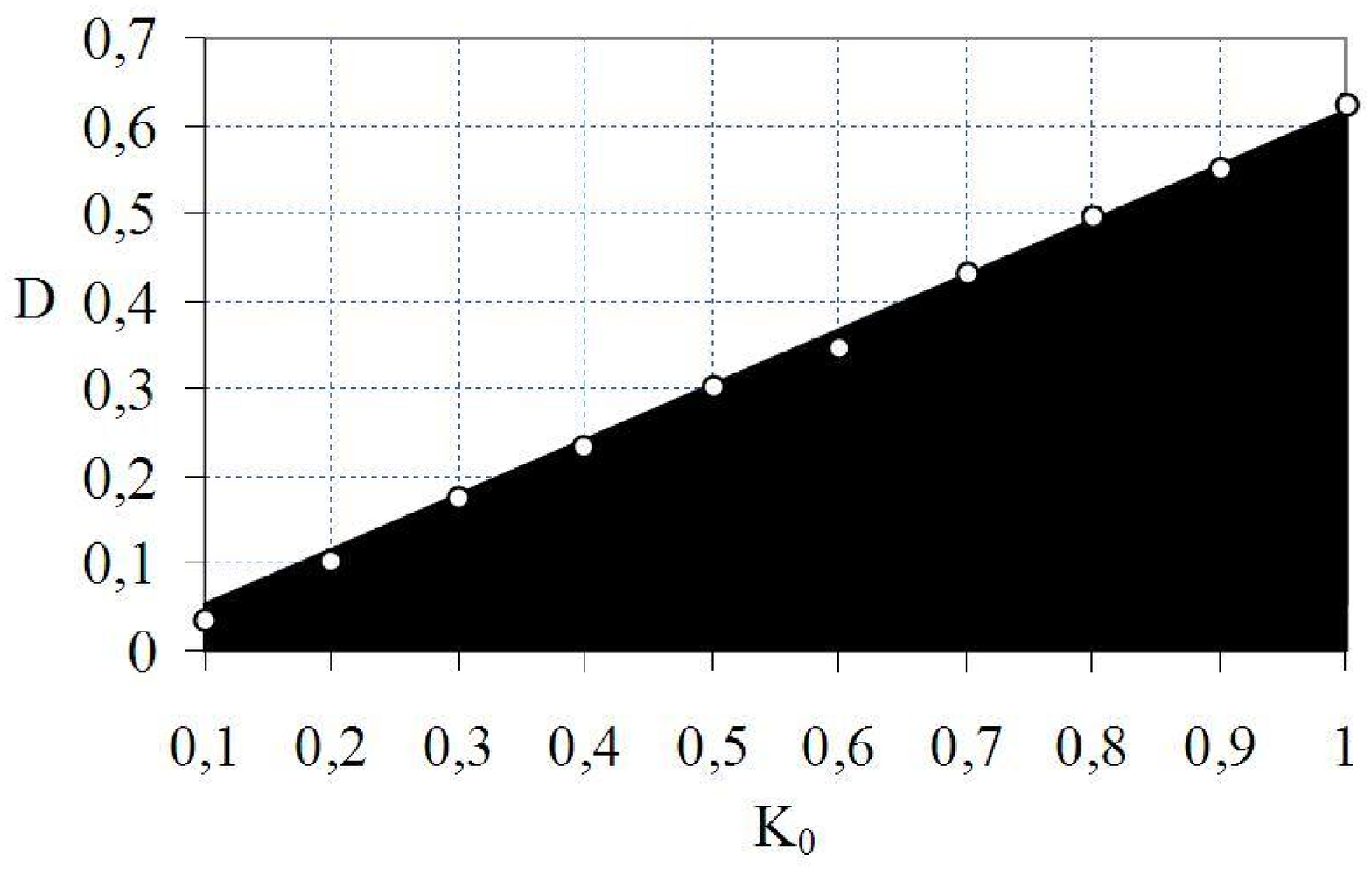,width=2.8in}}
\caption{Phase diagram in the plane $(K_0,D)$. The collapse region is in black, for $|J_2|=0.26 |J_1|$.  Other parameters are $D_1=D_2=a$, $I_0=0.5$, $\epsilon=1$.} \label{fig:KD}
\end{figure}

Now we consider the lattice with the presence of itinerant spins.  As far as the interaction between itinerant spins is attractive, we need a chemical potential to avoid the collapse of the system. The strength of the chemical potential $D$ depends on $K_0$. We show in Fig. \ref{fig:KD} the collapse phase diagram which allows to choose for a given $K_0$, an appropriate value of $D$.

\begin{figure}[th]
\vspace*{8pt}
\centerline{\epsfig{file=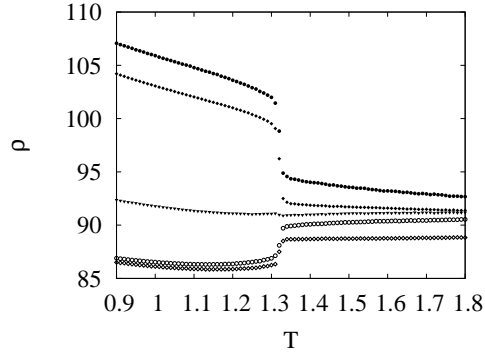,angle=0,width=2.8in}}
\caption{Spin resistivity versus $T$ for $|J_2|=0.26 |J_1|$ for several values of $D_1$: from up to down $D_1=0.7a$, $0.8a$, $0.94a$, $a$, $1.2a$.    Other parameters are  $N_x=N_y=20, N_z = 6$, $I_0=K_0=0.5$, $D_2=a$, $D=1$,  $\epsilon=1$.} \label{fig:R}
\end{figure}

We show now the main result on the spin resistivity versus $T$ for $|J_2|=0.26|J_1|$ for several values of $D_1$.  Other parameters are the same as in Fig. \ref{fig:ME}.  As said in section \ref{choice}, within the physical constraints, the variation of most of the parameters does not change qualitatively the physical effects observed in simulations, except for the parameter $D_1$. Due to the AF ordering, increasing $D_1$ means that we include successively
neighboring down and up spins surrounding a given itinerant spin.   As a consequence,  the energy of the itinerant spin oscillates with varying $D_1$, giving rise to the change of behavior of $\rho$:  $\rho$ can make a down fall or an upward jump at $T_C$ depending on the value of $D_1$ as shown in Fig. \ref{fig:R}. Note the discontinuity of $\rho$ at $T_C$.  This behavior has been observed in the frustrated FCC antiferromagnet.\cite{Magnin2}

\section{Concluding Remarks}

From the results shown above for the strongly frustrated $J_1-J_2$ model, we conclude that the spin resistivity reflects the nature of the first-order transition: it undergoes a discontinuity at the transition temperature. The fact that as $T\rightarrow 0$, $\rho$ increases slowly stems from the freezing of itinerant spins with decreasing $T$.  This has been experimentally observed  in ferromagnets and antiferromagnets as seen in  Fig. 11 of the paper by Chandra et al. on CdMnTe,\cite{Chandra}
 Fig. 2 of the paper by Du et al. for MnFeGe,\cite{Du} Fig. 6a of the paper by McGuire et al. on AF superconductors LaFeAsO,\cite{McGuire} Fig. 2 of the paper by Lu et al. on AF LaCaMnO,\cite{Lu} and Fig. 7 of the paper by Santos et al. on AF LaSrMnO.\cite{Santos}

We hope that these MC results will stimulate further theoretical calculations and experiments.

\end{document}